\documentclass[twocolumn,showpacs,preprintnumbers,amsmath,amssymb]{revtex4}

\usepackage{graphicx}
\usepackage{dcolumn}
\usepackage{bm}
\usepackage{amssymb}
\usepackage{indentfirst}
\usepackage{psfig,color}
\usepackage{epsfig}
\usepackage{epsf}
\usepackage{graphicx}
\usepackage{slashbox}

\begin{document}

\title{Hadrons in AdS/QCD correspondence.}
\author{Alfredo Vega} \author{Ivan Schmidt}
\affiliation{Departamento de F\'\i sica y Centro de Estudios
Subat\'omicos,  Universidad T\'ecnica Federico Santa Mar\'\i a,
Casilla 110-V, Valpara\'\i so, Chile}

\begin{abstract}

We present an holographical soft wall model which is able to
reproduce not only Regge spectra for hadrons with arbitrary integer
spin, but also with spin 1/2 and 3/2, and with an arbitrary number
of constituents. The model includes the anomalous dimension of
operators than create hadrons, together with a dilaton, whose form
is suggested by Einstein equations and the AdS metric used.

\end{abstract}

\preprint{USM-TH-248}

\maketitle

\section{Introduction.}

From its beginnings, progress in QCD at low energies has seen
impeded because there are no good tools available in order to work
with strongly coupled Yang Mills theories. Nevertheless, in the last
few years the AdS / CFT correspondence, or more generally the use of
Gauge / String dualities, has provided a new approach that could
improve this situation.

At present a dual to QCD is unknown, but a simple approach known as
Bottom - Up has been quite successful. In fact, this kind of models
gives us a way to deal with QCD that includes counting rules at
small distances and confinement at large distances, and has been
shown to be successful in several concrete QCD applications, such as
in hadronic scattering processes \cite{PolStrass1, Janik, BdT1,
Levin}, hadronic spectra \cite{BdT2, KKSS, Forkel, VegaSchmidt},
hadronic couplings and chiral symmetry breaking \cite{DaRol, EKSS, Colangelo},
quark potentials \cite{Boschi, Andreev1, Jugeau} and weak hadronic decays
\cite{Hambye}.

As was shown by Polchinski and Strassler \cite{PolStrass1}, it is
possible to simulate confinement introducing a cut off in the
holographic coordinate z. This kind of models is known as hard wall
models (HW), albeit phenomenologically they have problems, since the
obtained spectra does not have Regge behavior. To remedy this it is
necessary to introduce a soft cut off, using a dilaton field
\cite{KKSS, DaRol}, or using a warp factor in the metric
\cite{Andreev2, Ghoroku}. These models are called soft wall models
(SW).

In the literature it is possible to find holographic models applied
to glueball and mesons, which exhibit a linear dependence between
the hadronic mass squared and both the angular momenta and the
radial quantum number. However, this situation is different in the
baryonic sector, where Regge like spectra could be obtained
considering models without dilaton and using a warp factor in the
AdS metric \cite{Forkel}, or using integrability method for AdS /
CFT equations \cite{BdT0}. Apart from this, with the exception of
glueballs, exotic hadrons have not been considered in a general way,
except in Ref. \cite{VegaSchmidt}, where general scalar hadrons were
discussed.

This work aims on one side to extend the ideas presented in
\cite{VegaSchmidt}, in order to consider not only hadrons with
arbitrary integer spin, but also with spin 1/2 and 3/2, and with an
arbitrary number of constituents, in a SW model that exhibits Regge
behavior for these hadrons. Moreover, we include the anomalous
dimension of operators which create the hadrons that we consider.
For these purposes a crucial point is to consider a different
dilaton than the usual quadratic choice.

In the usual SW models, the dilaton is factorized from the AdS Dirac
equation, and therefore there is no advantage or improvement over
the HW situation in this case. As we will see soon, the inclusion of
the anomalous dimension allows us to improve this aspect.

To take into account anomalous dimensions is not only motivated by
the previous argument, since in the AdS / CFT context each operator
in the CFT side is related to an AdS mode in the bulk, according to
a specific dictionary. Nevertheless, in order to apply the
correspondence to a theory like QCD, it is important to consider
that with the exception of conserved currents, the operators have
dimensions that are scale dependent due to anomalous dimensions, and
this aspect should be included in models applied to QCD based in the
correspondence. This point was developed in \cite{Cherman}, where
the anomalous dimension introduced a z dependence in the mass of AdS
modes associated to operators, and in this way it could affect other
quantities that depend on this scale.

Another aspect that needs to be stressed is that these holographic
models consider QCD  in the conformal limit, where it is weakly
coupled due to asymptotic freedom. As a result, it is far from
obvious that the use of a classical, weakly curved 5D background is
justified, because you can expects a dual of QCD like a string
theory on some highly curved space (as noted for instance in
\cite{KKSS}). However, despite this possible problem, it is
interesting to try to investigate if a classical 5D background might
serve as phenomenological useful approximation to a holographical
dual of QCD, and many Bottom - Up models show a remarkable agreement
with experimental data and this work must be considered in this
perspective.

The present work has been structured as follow. Section II is
dedicated to hadrons with an integer arbitrary spin. In section II
we consider hadrons with spin 1/2, and in IV we show our results,
which include hadronic spectra and the pion form factor. Finally we
present in section V some conclusions.

\section{Hadrons with arbitrary integer spin.}

We begin by considering an asymptotically AdS space defined by the metric

\begin{equation}
 \label{metrica}
 ds^{2} = e^{2 A(z)} ( \eta_{\mu\nu} dx^{\mu} dx^{\nu} ) ,
\end{equation}
and the action for arbitrary integer spin modes is \cite{KKSS,HYY}

\begin{equation}
 I = \frac{1}{2} \int d^{5}x \sqrt{g} e^{-\Phi (z)} [ \triangle_{N} \phi_{M_{1}...M_{S}} \triangle^{N} \phi^{M_{1}...M_{S}} + \nonumber
\end{equation}
\begin{equation}
 m_{5}^{2} \phi_{M_{1}...M_{S}} \phi^{M_{1}...M_{S}} ] ,
\end{equation}
where $\Phi (z)$ is a dilaton field that only depends on the
holographical coordinate z.

From this action, the equation of motion for the part propagating in
the bulk can be written in general as \cite{HYY}

\begin{equation}
 \label{EcuacionBosonGeneral}
 \partial^{2} \varphi - (\partial B(z)) \partial \varphi + [ M^{2} - m_{5}^{2} e^{2 A(z)} ] \varphi = 0 ,
\end{equation}
Here the following equation was used

\begin{equation}
 B(z) = \Phi (z) - k (2 S - 1) A (z),
\end{equation}
where k is a constant and S corresponds to the spin of the mode
considered.

As was mentioned before, we introduce some modifications with
respect to the traditional SW models, where in general an AdS metric
($A(z) \sim - \ln (z)$) and a quadratic dilaton are considered. In
this work the dilaton that we use is suggested by Einstein's
equations \cite{Gursoy1, Gursoy2, DePaula}, which together with
other modifications to be made explicit a bit later, will allow us
to obtain a Regge type spectra.

Let us see how the specific form for the dilaton can be obtained. We
start with the action for 5D gravity coupled to a dilaton
\cite{Gursoy1, Gursoy2, DePaula}:
\begin{equation}
 S = \frac{1}{2 k^{2}}\int d^{5}x \sqrt{g} \biggl(-R - V(\Phi) + \frac{1}{2} g^{\mu \nu} \partial_{\mu}\Phi \partial_{\nu}\Phi \biggr),
\end{equation}
where k is Newton's constant in 5D and $V(\Phi)$ is the scalar field
potential. From equation (\ref{metrica}), we find the coupled
Einstein's equations
\begin{equation}
 \label{E1}
 6 A^{\backprime 2} - \frac{1}{2} \Phi^{\backprime 2} + e^{2 A(z)} V(\Phi) = 0 ,
\end{equation}
\begin{equation}
 \label{E2}
 3 A^{\backprime 2} + 3 A^{\backprime \backprime} + \frac{1}{2} \Phi^{\backprime 2} + e^{2 A(z)} V(\Phi) = 0,
\end{equation}
\begin{equation}
 \label{E3}
 \Phi^{\backprime 2} + 3 A^{\backprime} \Phi - e^{2 A(z)} \frac{d V}{d \Phi} = 0.
\end{equation}

Einstein's equations, (\ref{E1}) and (\ref{E2}), determine the
dilaton directly from the metric as:
\begin{equation}
 \Phi^{\backprime} = \sqrt{3 A^{\backprime 2} - 3 A^{\backprime \backprime}}.
\end{equation}

Our model is defined by

\begin{equation}
 A(z) = \rho \ln \biggl( \frac{R}{z} \biggr),
\end{equation}
where $\rho$ is a constant. With this choice
\begin{equation}
 \Phi(z) = \lambda ln(z),
\end{equation}
where $\lambda = \sqrt{3 \rho (\rho - 1)}$ and in order to get equations with exact solutions, we will use $\lambda = 2$.

With this, and considering $\beta = - k (2 S - 1)$, equation
(\ref{EcuacionBosonGeneral}) changes to

\begin{equation}
 \label{EcuacionBoson2.0}
 \partial^{2} \varphi - \biggl(\frac{\lambda}{z} - \frac{\beta}{z}\biggr) \partial \varphi + \biggl[ M^{2} - \frac{m_{5}^{2} R^{2}}{z^{2}} \biggr] \varphi = 0 .
\end{equation}

The hadronic spectra is obtained from this equation, but as was
shown for the scalar case in \cite{VegaSchmidt}, the form of
$m_{5}^{2} R^{2}$ must be obtained considering the AdS / CFT
dictionary, according to which, modes of the gravity theory and
physical states of the gauge theory satisfy an UV boundary condition
at $z \rightarrow 0$. More specifically, for a dual state $| i
\rangle$ with spin 0 one must select solutions with behavior
$\varphi (z) \rightarrow z^{\Delta}$ when $z \rightarrow 0$, where
$\Delta$ is the conformal dimension of the gravity mode, and which
must be equal to [O], the dimension of the operator that creates the
state $| i \rangle$ in the gauge theory. Modes with spin S need an
additional factor $z^{-S}$, and then the boundary condition is
generalized when $z \rightarrow 0$ and the dimension of the operator
[O] is changed to the respective twist $\tau$ \cite{PolStrass1,
PolStrass2}

\begin{table}[h]
\begin{center}
\begin{tabular}{ c c c  c c c  c c c }
  & $\varphi (z) \rightarrow z^{\tau}$ & & & and & & & $\tau = [O] - S$. & \\
\end{tabular}
\end{center}
\end{table}

Different values for the product $m_{5}^{2} R^{2}$ are related to
different kinds of hadrons, and for this reason it is necessary to
analyze the possible values for this product. This is achieved by
equating the conformal dimension ($\Delta$) of the mode propagating
in AdS with the dimension of the operator that creates hadrons
([O]). At this point we introduce additional modifications with
respect to traditional holographic models.

In the first place we consider the conformal dimension of the AdS
modes, which is extracted from the behavior of the solutions of
(\ref{EcuacionBoson2.0}) when $z \rightarrow 0$. Thus
\begin{equation}
 \label{DimConforme}
 \Delta = \frac{1}{2} (1 - \beta + \lambda) + \frac{1}{2} \sqrt{\beta^{2} + 4 m_{5}^{2} R^{2} - 2 \beta (1 + \lambda) + (1 + \lambda)^{2}}.
\end{equation}

We also need the dimension of the operator that creates hadrons,
where we have included a term that represents the anomalous
dimension of this kind of operators. In principle this dimension can
be written as

\begin{equation}
 [ O ] = \Delta_{0} + L + \delta,
\end{equation}
where $\Delta_{0}$ has contributions from quarks, antiquarks and/or
gluons, $L$ is a contribution coming from the angular momentum and
$\delta$ is the anomalous dimension.

The anomalous dimension is proportional to the coupling constant
\cite{Cherman}, which in turn is related in holographical models to
the dilaton, in the form $\sim e^{\Phi(z)}$ \cite{Gursoy1, Gursoy2}.
According to the correspondence, the holographical coordinate is
related to the energy of the theory that lives in the edge, and then
the anomalous dimension that we are considering is a quantity that
depends on energy. With this in mind, the $\tau$ for the operator
can be finally written as

\begin{equation}
 \label{DimMasa}
 \tau = \Delta_{0} + L + \omega e^{\Phi(z)} - S,
\end{equation}
where $\omega$ is a constant.

Now we have expressions for both $\Delta$ and $\tau$. Equating them
gives a result for to $m_{5}^{2} R^{2}$, which is

\begin{equation}
 m_{5}^{2} R^{2} = (\Delta_{0} + L - S + \omega z^{2})(\Delta_{0} + L - S - 3 + \beta + \omega z^{2}),
\end{equation}
where we took $\lambda = 2$. This choice allows us to get analytical
solutions with Regge behavior from equation
(\ref{EcuacionBoson2.0}).

Using this in (\ref{EcuacionBoson2.0}), the normalizable solutions
are

\begin{equation}
 \label{ModoBoson}
 \varphi (z) = C e^{-\frac{1}{2} \omega z^{2}} z^{\Delta_{0} + L - S} L_{n}^{m} (\omega z^{2}),
\end{equation}
where $L_{n}^{m} (x)$ are Laguerre polynomials, and

\begin{equation}
 m = - 1 + \frac{1}{2} (-1 + \beta + 2 L - 2 S + 2 \Delta_{0}), \nonumber
\end{equation}
\begin{equation}
 n = \frac{M^{2} + 2 \omega}{4 \omega} - [-1 + \frac{1}{2}(-1 + \beta + 2 L - 2 S + 2 \Delta_{0})] - 1 \nonumber \
\end{equation}
\begin{equation}
 n  = 0, 1, 2, ... \nonumber
\end{equation}
From the last equation we get the spectrum, which is

\begin{equation}
 \label{EspectroEspinEntero}
 M^{2} = 4 \omega \biggl[n + L + \biggl(\Delta_{0} + \frac{\beta}{2} - 1 - S \biggr)\biggr].
\end{equation}

In (\ref{ModoBoson}) C is a normalization constant, which can be
fixed using

\begin{equation}
 \label{Normalizacion}
 \int_{0}^{\infty} dz z^{\beta} e^{-\Phi(z)} | \varphi(z) |^{2} =  1 .
\end{equation}

\section{Spin 1/2 Hadrons.}

As was said before, the dilaton field can be factorized from the
Dirac AdS equation, and for this reason this field does not have any
influence over the spectrum in a traditional SW model. Nevertheless,
since in our model we consider anomalous dimensions, $m_{5}$ has a
dependence on the dilaton, and in this way this field can affect the
spectrum. This allows us to get the desired Regge behavior in this
sector, as we will see soon.

According to \cite{Kirsch} the Dirac equation with dilaton in AdS
space can be written as

\begin{equation}
 \biggl( \displaystyle{\not} D - \frac{1}{2} e^{M}_{A} \gamma^{A} \partial_{M} \Phi - m \biggr) \Psi (x^{\mu},z) = 0 ,\nonumber
\end{equation}
where $\displaystyle{\not} D = e^{M}_{A} \gamma^{A} D_{M} $ and
$e^{M}_{A}$ correspond to a $(d+1)$-bein.

Dual modes to baryons can be decomposed into left and a right pieces
\cite{Forkel}

\begin{equation}
\Psi (x^{\mu},z) = \biggl[ \frac{1+\gamma^{5}}{2} f_{+}(z)
+\frac{1-\gamma^{5}}{2} f_{-}(z) \biggr] \Psi_{4}(x^{\mu}) ,\nonumber
\end{equation}
where $\Psi_{4}(x^{\mu})$ satisfies the Dirac equation in four
dimensions.

Following the procedure described in \cite{Kirsch}, the dilaton
field can be factorized, and applying $\displaystyle{\not} D$ over
the Dirac equation and using $\gamma^{5} f_{\pm} = \pm f_{\pm}$, we
get an equation for the part that is propagating in the bulk, which
is

\begin{equation}
 \label{EcuacionDirac}
 \partial^{2} f_{\pm} - \frac{4}{z} \partial f_{\pm} + \biggl[ M^{2} + \frac{6}{z^{2}} - \frac{m_{5}^{2} R^{2}}{z^{2}} + \frac{\gamma m_{5} R}{z^{2}} \biggr] f_{\pm} = 0 ,
\end{equation}
where $\gamma = \pm 1$, depending on whether we are considering the
left or right part.

This equation is the same that one gets in HW models, with solutions
that are Bessel functions and whose spectrum in this case does not
have Regge behavior \cite{BdT2}. The last point is true when $m_{5}
R$ is independent on z.

Considering the solutions for (\ref{EcuacionDirac}) it is possible
to see that the conformal dimension is

\[ \Delta = \left\lbrace
  \begin{array}{c l}
    \frac{5}{2} + \frac{1}{2} |1 - m_{5} R| & \text{if $\gamma = 1$}\\
    3 + m_{5} R & \text{if $\gamma = -1$}.
  \end{array}
\right. \]

Equating this with (\ref{DimMasa}) a specific form for $m_{5} R$ is
obtained, which replaced in (\ref{EcuacionDirac}) enables us to
obtain a spectrum that can be written as

\begin{equation}
 \label{EspectroDirac}
 M^{2} = 4 \omega \biggl[n + L + \biggl(\Delta_{0} - \frac{5}{2} \biggr)\biggr].
\end{equation}
Here, just as in the integer spin case and for the same reasons, we
have taken $\lambda = 2$. In this way, we have got the same dilaton
for all cases considered.

It should be noted that $\gamma = 1$ actually represents two cases.
When $2 m_{5} R > 1$ the spectrum is given by (\ref{EspectroDirac}),
which is the result that appears when $\gamma = -1$. On the other
hand, the other solution that corresponds to the case $\gamma = 1$ gives
an $M^{2}$ which is negative.

\section{Phenomenological implications.}

\subsection{Hadronic spectrum.}

In the model that we have presented the general form of the
spectrum, for all cases considered, is

\begin{equation}
 \label{Masas}
 M^{2} = A [n + L + v],
\end{equation}
where $A = 4 \omega$ is the Regge slope, and $v$ is given by

\[ v = \left\lbrace
  \begin{array}{c l}
    \Delta_{0} + \frac{\beta}{2} - 1 - S & \text{; If S is integer}\\
    \Delta_{0} - \frac{5}{2} & \text{; spin 1/2}.
  \end{array}
\right. \]

Notice the the model give us the Regge slope in terms of $\omega$,
but it does not provide information on how to calculate it, and
therefore this constitute a phenomenological input in our model.

(\ref{Masas}) can be applied to different kinds of hadrons with an
arbitrary number of constituents. Unlike the model presented in
\cite{VegaSchmidt}, where v was a parameter that had to be adjusted
for each kind of hadron, and which restricted the predictive power
of the model, here we can see that for the integer spin case $v$
depends on $\Delta_{0}$ and $\beta$, where $\Delta_{0}$ is obtained
from the number of quark, antiquarks and gluons in the hadron
considered and moreover $\beta$ depends on the spin, taking a single
value for scalars, another value for vectors and so on. After
adjusting this parameter using data for a specific hadron, it can be
used for other hadrons with the same spin. On the other hand, in the
spin 1/2 case $v$ depends only on the number of quarks, antiquarks
and gluons in the hadron considered.

Before going into some specific examples, it is relevant to make a
brief comment about the Regge slope. The Regge structure for the
spectrum is a very good approximation for mesons and light baryonic
resonances, and although the adjustment of the slope gives different
values for different hadrons, its value changes very little, and
therefore with good approximation can be considered universal.

Two adjustments to meson data that produce two mutually consistent
slopes give  $A = 1.25 \pm  0.15 GeV^{2}$ \cite{Anisovich} ($\omega \sim 0.313
GeV^{2}$) and $A =
1.14 \pm 0.013 GeV^{2}$ \cite{Bugg} ($\omega \sim 0.285
GeV^{2}$). Adjusting to light baryonic
resonances (i.e those formed by quarks u, d and s) gives $A = 1.081
\pm  0.035 GeV^{2}$ \cite{Klempt} ($\omega \sim 0.270
GeV^{2}$). From this, the value $A \sim 1.1
GeV^{2}$ ($\omega \sim 0.275
GeV^{2}$) can be considered approximately universal for all
trajectories \cite{Iachello}.

Note that according the dilaton used ($\Phi(z) = 2 ln (z)$), the dimensionless anomalous dimension is $\delta = \omega e^{\Phi(z)} = \omega z^{2}$. With this is clear that $\omega$ units must be square of Energy.

\subsubsection{Scalar hadronic spectrum.}

As was mentioned before, we take the value 1.1$Gev^{2}$ for the
Regge slope ($\omega \sim 0.275 GeV^{2}$). We must also fix $\beta$, and for this we consider the
pion form factor, which as we will see below in section IV.B, is
given by (\ref{FactorForma}). This, when $Q = 0,$ must be reduced to
the normalization condition (\ref{Normalizacion}), and then $\beta =
-3$, which makes the normalization to be the same as the one that is
used in \cite{BdT3}.

The spectrum for scalar mesons is shown in Fig 1, while some
examples about model prediction for scalar exotic hadrons appear in
Table 1. This includes the content of quarks, antiquarks (that
contribute with 3/2 to $\Delta_{0}$) and gluons (that contribute
with 2 to $\Delta_{0}$).

\begin{figure}[h]
  \begin{tabular}{cc}
    \includegraphics[width=1.7in]{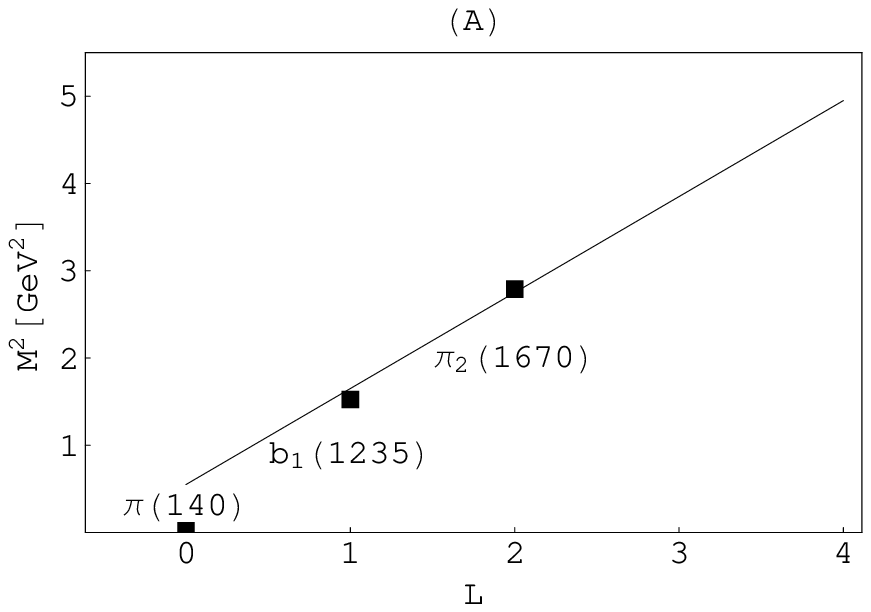} &
    \includegraphics[width=1.7in]{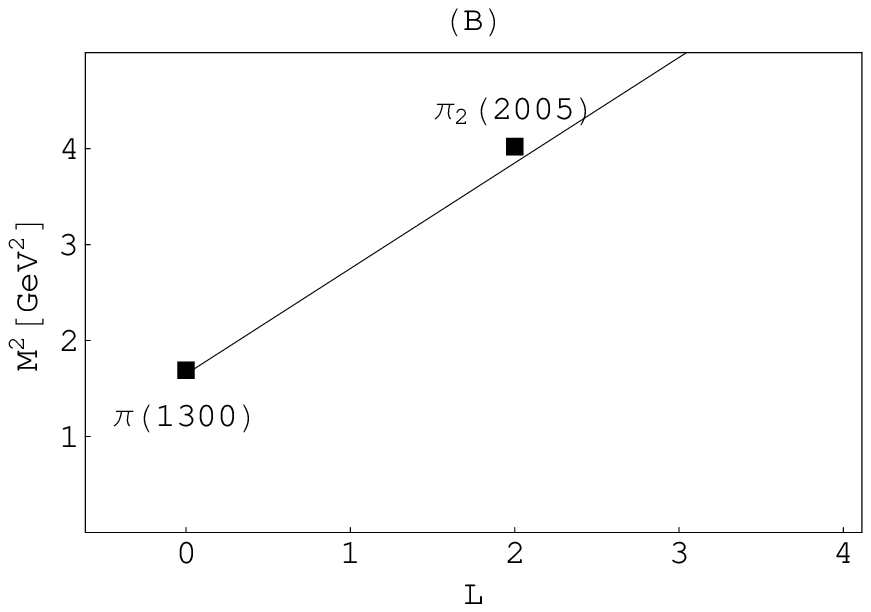}
  \end{tabular}
  \begin{tabular}{cc}
    \includegraphics[width=1.7in]{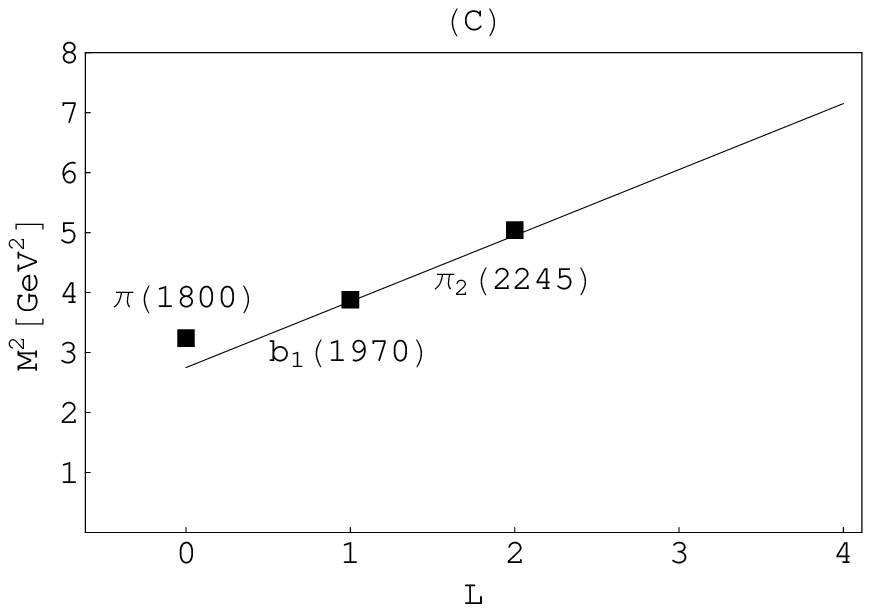} &
    \includegraphics[width=1.7in]{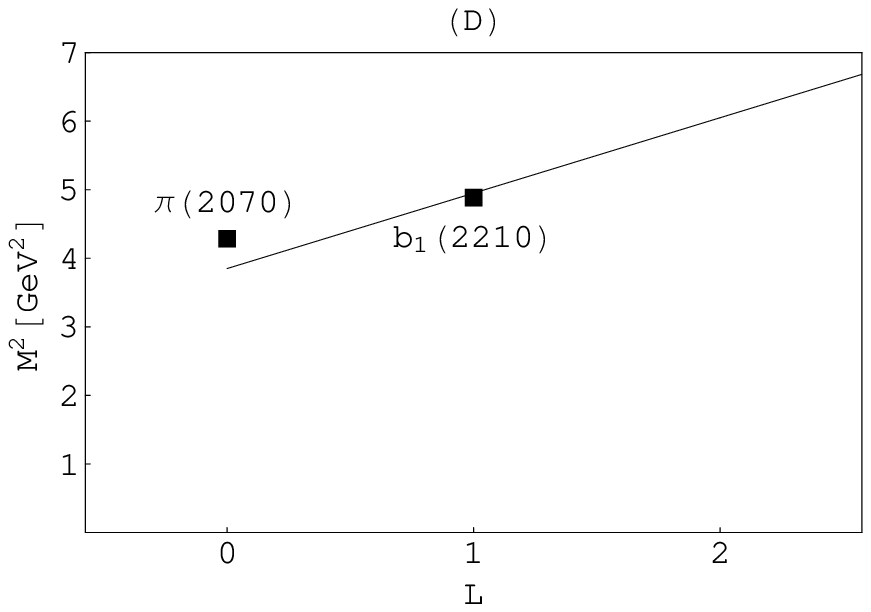}
  \end{tabular}
\caption{Spectra of scalar mesons, calculated within the Soft Wall
model. The figures correspond to different radial excitations. (A)
$n=0$ (B) $n=1$ (C) $n=2$ (D) and $n=3$. }
\end{figure}

\begin{table}[h]
\begin{center}
\caption{Scalar exotic hadron masses, with $n = L = 0$. We consider
hadrons with n quarks (and / or antiquarks) and m gluons.\\}
\begin{tabular}{ c c c | c c c | c c c }
  \hline
  \hline
  & $\Delta_0$ & & & (nQ)(mG) & & & M [GeV] & \\
  \hline
  & 4 & & & (2G) & & & 1.28 & \\
  & 5 & & & (2Q)(1G) & & & 1.66 & \\
  & 6 & & & (4Q) & & & 1.96 & \\
  & 7 & & & (2Q)(2G) & & & 2.22 & \\
  & 8 & & & (4Q)(1G) ; (4G) & & & 2.46 & \\
  & 9 & & & (6Q) ; (2Q)(3G) & & & 2.67 & \\
  & 10 & & & (4Q)(2G) & & & 2.87 & \\
  \hline
  \hline
\end{tabular}
\end{center}
\end{table}

\subsubsection{Vector hadron spectrum.}

In this case we proceed in a similar way as with the scalar case,
and we get $\beta = -1$. With this the normalization is the same as
the one used in \cite{GrigoRadi}, and the Regge slope is the same as
the one used above for the scalar case.

The spectrum for scalar mesons is shown in Fig 2, while some
examples about model prediction for scalar exotic hadrons appear in
Table II.

\begin{figure}[h]
  \begin{tabular}{cc}
    \includegraphics[width=1.7in]{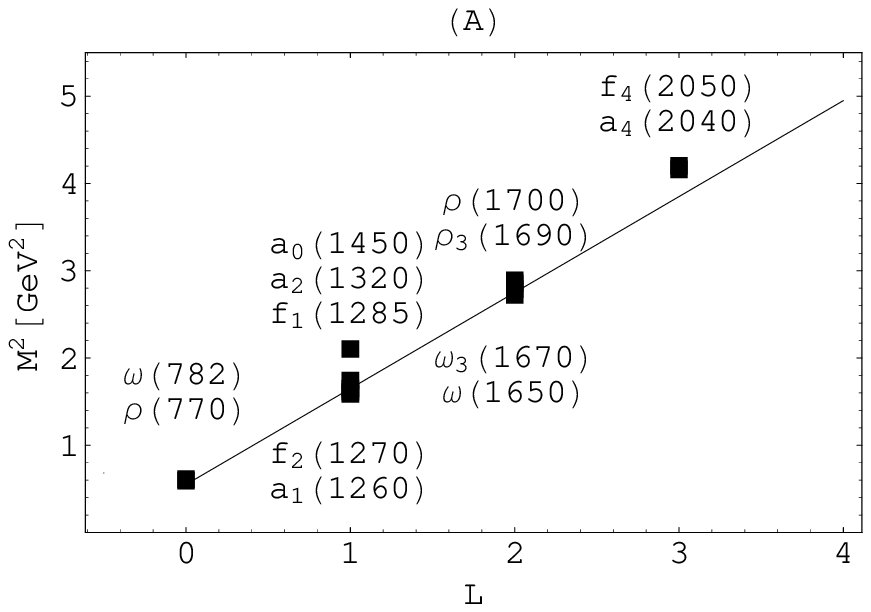} &
    \includegraphics[width=1.7in]{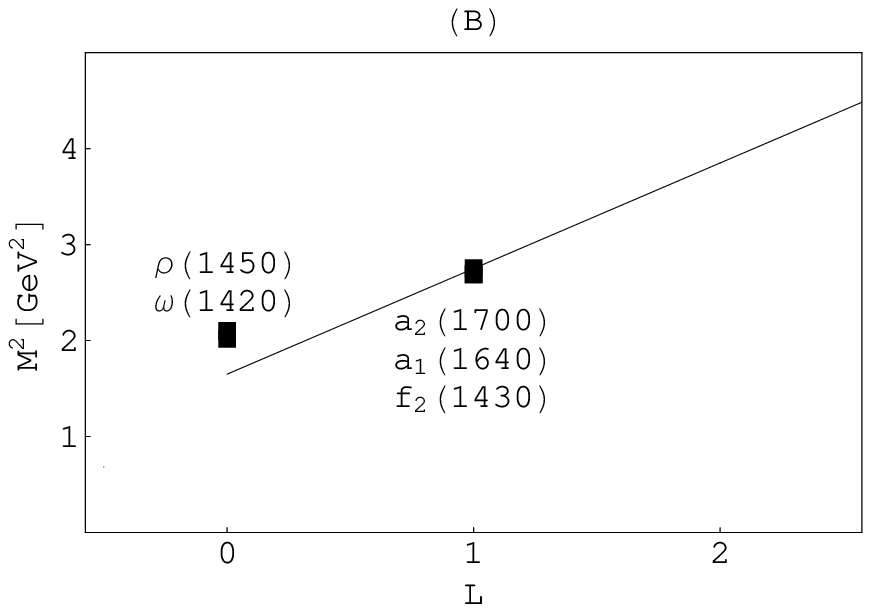}
  \end{tabular}
  \begin{tabular}{cc}
    \includegraphics[width=1.7in]{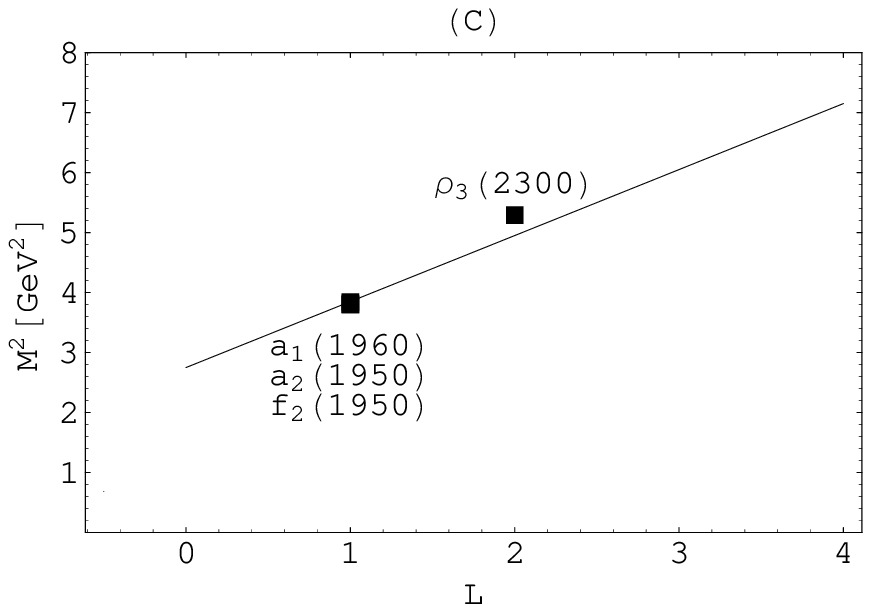} &
    \includegraphics[width=1.7in]{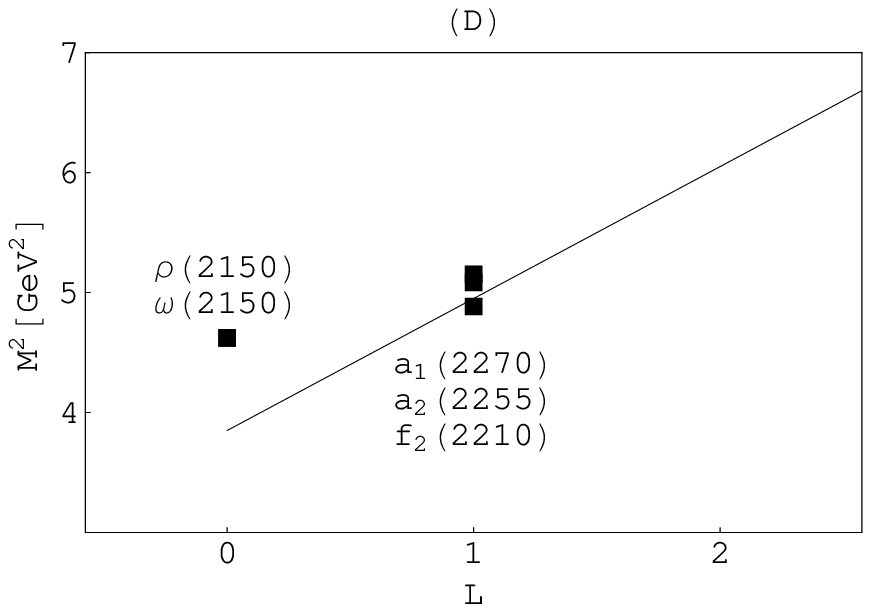}
  \end{tabular}
\caption{Spectra of vector mesons, calculated within the Soft Wall
model. The figures correspond to different radial excitations. (A)
$n=0$ (B) $n=1$ (C) $n=2$ (D) and $n=3$. }
\end{figure}

\begin{table}[h]
\begin{center}
\caption{Exotic vector hadron masses, with $n = L = 0$. We consider
hadrons with n quarks (and / or antiquarks) and m gluons.}
\begin{tabular}{ c c c | c c c | c c c }
  \hline
  \hline
  & $\Delta_0$ & & & (nQ)(mG) & & & M [GeV] & \\
  \hline
  & 5 & & & (2Q)(1G) & & & 1.66 & \\
  & 6 & & & (4Q) ; (3G) & & & 1.96 & \\
  & 7 & & & (2Q)(2G) & & & 2.22 & \\
  & 8 & & & (4Q)(1G) & & & 2.46 & \\
  & 9 & & & (6Q) ; (2Q)(3G) & & & 2.67 & \\
  & 10 & & & (5G) ; (4Q)(2G) & & & 2.87 & \\
  \hline
  \hline
\end{tabular}
\end{center}
\end{table}

\subsubsection{Spin 1/2 hadron spectrum.}

In this case $v$ does not need to be fixed from experimental data.
As one can see from Fig 3, using an universal Regge slope gives
results somewhat higher than the experiments, but using a value of
0.9 $[GeV^{2}]$ \cite{Forkel} ($\omega \sim 0.225
GeV^{2}$), adjusted to baryonic data, the
results are better.

Both values are used in Table 3, where model predictions for some
exotic spin 1/2 hadrons are shown.

\begin{figure}[h]
  \begin{tabular}{cc}
    \includegraphics[width=1.7in]{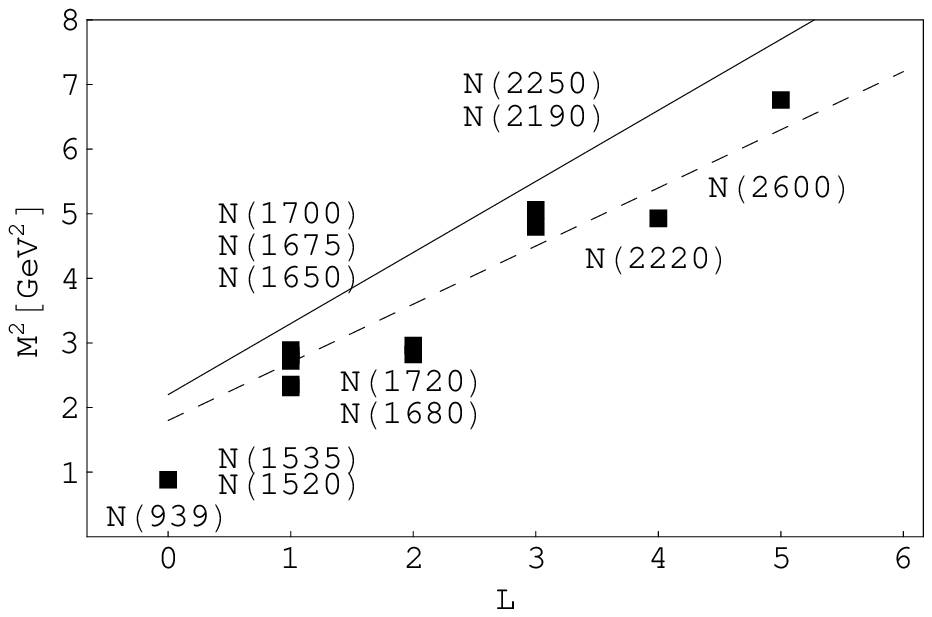} &
    \includegraphics[width=1.7in]{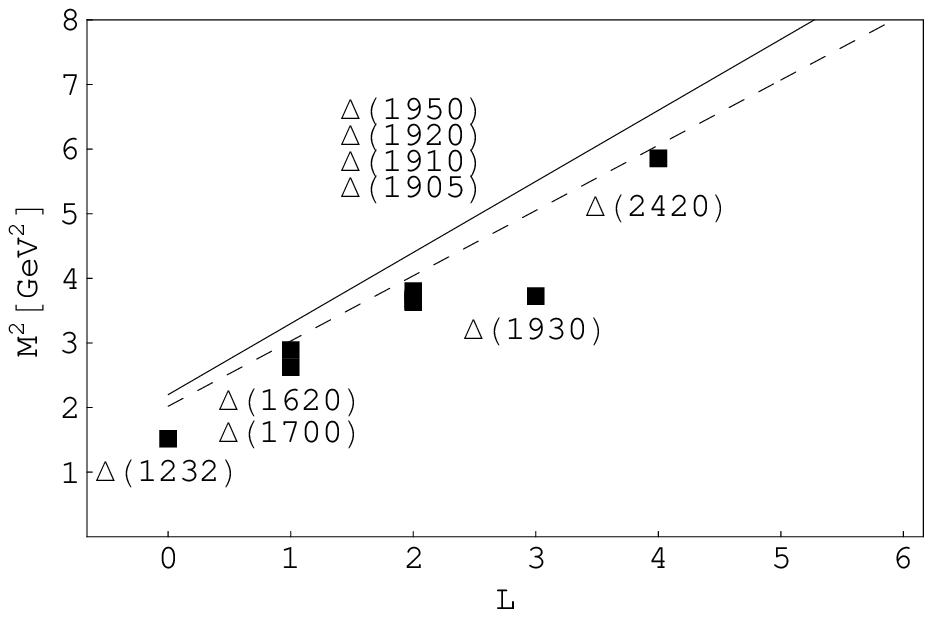}
  \end{tabular}
\caption{Nucleons and $\Delta$ resonances spectrum. The continuous
line is the model prediction using an  universal value of $A = 1.1
GeV^{2}$ ($\omega \sim 0.275
GeV^{2}$), while the dashed line was obtained using Regge slopes
adjusted to each case, with values $A = 0.9 GeV^{2}$ ($\omega \sim 0.225
GeV^{2}$) for nucleons
and $A = 1.01 GeV^{2}$ ($\omega \sim 0.253
GeV^{2}$) for $\Delta$ resonances \cite{Forkel}.}
\end{figure}

\begin{table}[h]
\begin{center}
\caption{Spin 1/2 exotic hadron masses with $n = L = 0$. We consider
hadrons with n quarks (and / or antiquarks) and m gluons. Column
$M_{U}$ was calculated using $A = 1.1 GeV^{2}$ ($\omega \sim 0.275
GeV^{2}$), the universal Regge
slope used in this work, while M contains the results obtained using
$A = 0.9 GeV^{2}$ ($\omega \sim 0.225
GeV^{2}$), a value fixed from nucleon data \cite{Forkel}.}
\begin{tabular}{ c c c | c c c | c c c | c c c }
  \hline
  \hline
  & $\Delta_0$ & & & (nQ)(mG) & & & $M_{U}$ [GeV] & & & M [GeV] & \\
  \hline
  & 13/2 & & & (1Q)(3G) & & & 2.10 & & & 2.01 & \\
  & 15/2 & & & (5Q) & & & 2.35 & & & 2.25 & \\
  & 17/2 & & & (3Q)(2G) & & & 2.57 & & & 2.46 & \\
  & 19/2 & & & (5Q)(1G) & & & 2.77 & & & 2.66 & \\
  & 21/2 & & & (3Q)(3G) ; (7Q) & & & 2.97 & & & 2.84 & \\
  & 23/2 & & & (5Q)(2G) & & & 3.15 & & & 3.01 & \\
  \hline
  \hline
\end{tabular}
\end{center}
\end{table}

\subsubsection{Spin 3/2 hadrons spectrum.}

Solutions to Rarita - Schwinger equation in AdS space are more
complex to get, but its spectrum is similar to the Dirac case
\cite{BdT2, Forkel}. As is possible to see in Fig 3, again the
results are somewhat high, but using $A = 1.01 GeV^{2}$ ($\omega \sim 0.253
GeV^{2}$), adjusted to
$\Delta$ resonances gives better results.

\subsection{Pion Form Factor.}

The model described in this work allows to get hadronic spectra and
the holographical mode associated to hadrons, with which it is
possible calculate form factors in AdS. In this section we are
interested in showing that the model, following  \cite{BdT3}, can be
used in aspects of hadronic physics that go beyond the spectrum
reproduction. Specifically we consider the pion electromagnetic form
factor as an example, calculated in the AdS side.

This application should be considered with some degree of caution,
since the pion is not really inside Regge trajectories. Then if one
takes the mode with $n = 0$ and $l = 0$ with the same value for
$\omega$ that was used for the Regge slopes,  a good holographical
description for the pion is not achieved. For this reason we
consider a different value for $\omega$ in the pion case.

The form factor in AdS is represented by an overlap integral in the
holographical coordinate of the normalizable modes, dual to incoming
and outgoing hadrons, $\varphi_{p}$ and $\varphi_{p^{,}}$, with the
non-normalizable mode $J(Q^{2}, z)$, dual to the electromagnetic
source propagating inside to AdS space \cite{PolStrass2, BdT3,
Grigoryan}, and which can be written as

\begin{equation}
 \label{FactorForma}
 F (Q^{2}) = \int_{0}^{\infty} \frac{dz}{z^{3}} e^{-\Phi (z)} J(Q^{2},z) | \varphi (z) |^{2}.
\end{equation}

Since the non-normalizable mode couples to the dilaton field,
$J(Q^{2},z)$ is obtained from the solution to
(\ref{EcuacionBoson2.0}), with $m_{5} = 0$, $P^{2} = - Q^{2}$,
$\beta = -1$ and $\lambda = 2$, and then

\begin{equation}
 J(Q^{2},z) = \frac{1}{2} z^{2} Q^{2} K_{2} (Q z),
\end{equation}
which is equal to 1 when either Q or z go to zero.

For the pion we consider equation (\ref{ModoBoson}) with $\Delta_{0}
= 3$, $n = 0$, $L = 0$ and $S = 0$, and then the mode that describes
this hadron is

\begin{equation}
 \varphi (z) = C e^{-\frac{1}{2} \omega z^{2}} z^{3},
\end{equation}
where C is a normalization constant, which is fixed by the
normalization (\ref{Normalizacion}).

\begin{figure}[h]
    \includegraphics[width=3.0in]{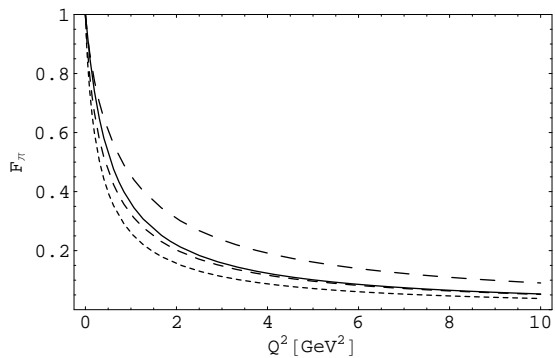}
\caption{Pion form factor. The continuous line corresponds to the
result obtained in \cite{BdT3}, while the dashed lines were obtained
using the model proposed here, for different values of $\omega$. The
lower dashed line uses $\omega = 0.05 GeV^{2}$, the intermediate dashed
line $\omega = 0.07 GeV^{2}$ and the upper dashed line is for $\omega =
0.13 GeV^{2}$.}
\end{figure}

\begin{figure}[h]
    \includegraphics[width=3.0in]{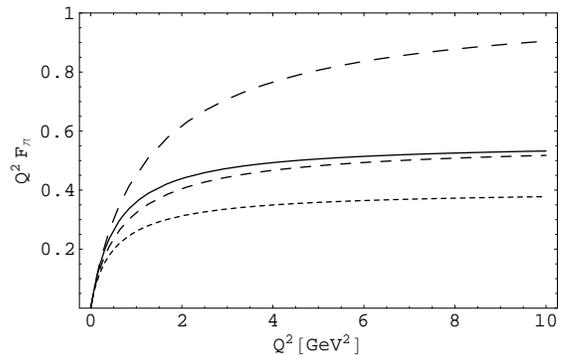}
\caption{$Q^{2} F_{\pi} (Q^{2})$. The continuous and dashed lines
correspond to the same cases considered in the previous graph.}
\end{figure}

In Figs. 4 and 5 we show graphs for $F(Q^{2})$ and $Q^{2} F(Q^{2})$,
where the continuous line corresponds to the result that appears in
\cite{BdT3}, while the dashed lines are results obtained with the
present model, for different values for $\omega$.

It is important to stress that in this work and in \cite{BdT3}, the
parameters used for the pion form factor are different from those
obtained with Regge trajectories, and this comes from the fact that
pions are an exception that does not fall into these trajectories.
In fact, in \cite{BdT3} the Regge slope is $4 \kappa^{2}$, and the
value of $\kappa$ used to graph $F_{\pi} (Q^{2})$ and $Q^{2} F_{\pi}
(Q^{2})$ is 0.375 GeV. With this value the Regge slope is almost
half of the phenomenological value. Pions are therefore an
exception, whose parameters have to be fixed in a different way.

Using the obtained pion form factor it is possible to extract the
mean square radius for this meson.

\begin{equation}
 \langle r_{\pi}^{2} \rangle = - 6 \frac{d F_{\pi} (Q^{2})}{d Q^{2}} |_{Q^{2} = 0} = \frac{3}{2 \omega},
\end{equation}
This is similar to the expression found in \cite{BdT3}. Considering
the values of $\omega$ used in Figs 4 and 5 for the results shown in
the dashed lines, the values for $\langle r_{\pi}^{2} \rangle$ are
$1.17 [fm^{2}]$, $0.83 [fm^{2}]$ and $0.45 [fm^{2}]$. The last value
corresponds to the experimental result, here obtained using $\omega
= 0.13 GeV^{2}$, which in turn corresponds to the upper dashed line in
Figs 4 and 5.

\section{Conclusions.}

In the present paper a SW model has been presented, which contains
some features that solve difficulties that appear frequently in
other models, and that are usually cause for criticism.

First, the model let us obtain hadronic spectrum with Regge
behavior, not only for the integer spin case, but also for spin 1/2
and 3/2. In order to do this we considered only one metric and one
dilaton, unlike what happens in Ref. \cite{Forkel}, where a family
of metrics was used. Secondly, we considered the anomalous dimension
for operators that create hadrons. And thirdly, the dilaton used has
a form that was suggested by Einstein's equations, corresponding to
the AdS metric that is used \cite{Gursoy1, Gursoy2, DePaula}. The
latter two traits allowed the model to reproduce Regge spectra in
all cases considered, and therefore the model can describe baryons
in a unified phenomenological model.

In \cite{VegaSchmidt} it was shown how to include hadrons with an
arbitrary number of constituents, although this was done only  for
the scalar case. Here this idea has been successfully extended to
other hadrons, presenting results for exotic hadrons with arbitrary
integer spin, and for spin 1/2 and 3/2. This was possible due to
that unlike in \cite{VegaSchmidt}, where $v$ in (\ref{Masas}) was a
parameter that needed to be fixed for each hadron, here we have two
situations. In the integer spin case $v$ depends on $\Delta_{0}$,
and through this it depends on the constituent number (but this is
easy to find) and on $\beta$, which depends on spin. Then $\beta$ needs
be fixed only one time for scalars, one time for vectors, and so on.
On the other hand, for the spin 1/2 and 3/2 cases, $v$ depends only
on $\Delta_{0}$. For this reason this model is really predictive in
the exotic hadronic sector, because the parameters A and $v$ in
(\ref{Masas}) could be fixed experimentally in some cases and are
calculable in others. This did not happen in \cite{VegaSchmidt}.

An interesting additional fact is the introduction of the anomalous
dimension, because this makes $m_{5}$ to be z dependent, and this
could have an effect on other quantities that depend on scales, like
chiral condensates and quarks masses, treated in holographic models
\cite{Cherman}.

Finally, for the pion form factor it is possible to adjust
parameters in a different way, because the pion does not fit into
Regge trajectories, and this requires giving it a special treatment
at the moment of fixing parameters.

\begin{acknowledgments}
A.V work was partially supported by HELEN program from CERN. A. V is
grateful to Nuria Rius and IFIC from Valencia, where part of this
work was done.

\end{acknowledgments}

\end{document}